# Sensitivity Analysis of calibration methods and factors effecting the statistical nature of radiation measurement

Mayank Goswami[1#], Kajal Kumari[1]
[1]Divyadrishti Laboratory, Department of Physics, IIT Roorkee, Roorkee, India
Department of Physics, IIT Roorkee, Roorkee, India
[#]mayank.goswami@ph.iitr.ac.in

## Abstract:

Scintillator detector's electronics is recalibrated against the datasheet given by the manufacturer. Optimal and mutual dependent values of (a) high voltage at PMT (Photomultiplier Tube), (b) amplifier gain, (c) average time to count the radiation particles (set by operator), and (d) number of instances/sample number are estimated. Total 5: two versions of Central Limit Theorem (CLT), (3) industry preferred Pulse Width Saturation, (4) calibration based on MPPC coupled Gamma-ray detector and (5) gross method are used. It is shown that CLT method is the most optimal method to calibrate detector and its respective electronics couple. An inverse modeling-based Computerized Tomography method is used for verification. It is shown that statistically averaging results are more accurate and precise data than mode and median, if the data is not skewed and random number of samples are used during the calibration process. It is also shown that the average time to count the radiation particle is the most important parameter affecting the optimal calibration setting for precision and accurate measurements of gamma radiation.

## 1. Introduction:

Scintillator detectors are preferred for a certain class of ionization radiation measurements (Baccouche et al., 2012; Chuong et al., 2019; Cuesta et al., 2013). The setup comprised of associated electronics requiring an operator to set several parameters(Freitas et al., 2015). For example, (a) gain value of amplifier, (b) range of discriminator/energy window, and (c) high voltage of photomultiplier tube (cathode to anode). Generally, the calibration of the typical system is performed onsite and the range of some of these settings is given a-priori (Canberra, 2000; Dimensions, 2009). These setting are widely considered stable, however, may require recalibration as the equipment completes its operational age, employed for different application or exposed to severe conditions (Beylin et al., 2005; Cabrera et al., 2000; Derenzo et al., 2014; Grinev et al., 1991). Besides using calibrated settings there are several other post-processing-related questions which are needed to be considered before measurement. These points usually are left for the operator to decide and may affect the outcome of the experiment, person to person(Prekeges, 2014). The calibration process itself requires finding the answer of the following measurement related questions:

1. How many times/instances one should repeat an experiment (the sample size $M$)?
2. How long one should carry out a single instance of measurement (Time $T$)?
3. What value one should take, average, median, or mode values of samples?
4. What should be the optimal setting of involved electronics?

Recording multiple instances (for arbitrary times) of the same event is usual practice to achieve better statistical inference. Generally, the mean value of all recorded values is used assuming that averaging of samples would not influence the measurement statistics (Prekeges, 2012).

Mode is rarely used as multiple local maxima may exist. Median values may be preferred (on case to case basis) over mean (Holt and Scariano, 2009), if data is skewed. For the small size of samples percentile bootstrap bias correction may improve the representation for the bulk of the observations using the median. Hierarchical shift functions are another class of tool as an alternative (HOCHBERG, 1988).

The sample size or instance of multiple recording, if are subthreshold or small may affect the reproducibility of the measurement (Albers and Lakens, 2018; Button et al., 2013; Colquhoun, 2014; Forstmeier et al., 2017).

If a measurement is done for a discrete variable, data from multiple instances is fitted to binary (if the nature of the variable is binary) or Poisson (Tsoulfanidis, 2010). However, several real-life variables, random in nature, are continuous. The preferred distribution in the field of measurement is the normal distribution, given in Eq. 1.

$$f(x) = \frac{1}{\sigma\sqrt{2\pi}} e^{-\frac{1}{2}\left(\frac{x-\bar{x}}{\sigma}\right)^2} \qquad (1)$$

Where $\bar{x}$ is the mean of the distribution and $\sigma$ is the standard deviation.

Statistically, if series of variables $x_i|_{i=1,\ldots,N}$ are repeatedly measured $M$ times, the average values $\bar{x}_i|_{i=1,\ldots,N}$ follow a normal distribution with a certain degree (2 or at best 3 sigma) of standard deviation. The impractical constraint $M \in (-\infty \text{ to } \infty)$ get normalized by a limited arbitrary time. The resemblance may improve with as many instances, as allowed, keeping practical limits under control. A widely used approximation is that the Binomial or Poisson distribution of a repeated random variable resembles Gaussian distribution if $M > 20$ (Tsoulfanidis, 2010).

The electronic settings of the radiation count/ events measurement device have a strong influence on the shape of pulses and resolution of the detector. B. Almutairi et al performed an experiment using the Geiger Muller counter to show the dependency of applied voltage on pulse shape (Almutairi et al., 2019).

The most widely used methods to estimate the optimal voltage and energy discriminator setting of a photo-multiplier tube (PMT) of scintillator detector are the gross method ("Method for Accurate Energy Calibration of Gamma Radiation Detectors with only Count Rate Data. (Conference) | OSTI.GOV," n.d.; Wang, 2003) and pulse width saturation/ plateau method (Almutairi et al., 2019; Marshall, 2014).

Gaussian nature of measurement noise often corrupts the data imparting non-negligible standard deviation. Central limit theorem (CLT) is another tool to estimate the relative level of precision and accuracy (Kumar et al., 2020; N. Tsoulfanidis, 2010).

## 1.1. Motivation

The information such as the number of samples $M$, how long each sample was taken $T$, if mean, mode or median was used to get a single value, remains absent from the datasheet for respective calibration settings. *We recalibrated the scintillator detector electronics to explore, if above mentioned statistical factors affect the suggested calibration settings.*

Five different methods ($N = 5$), namely: (a) two different implementations of CLT, (b) pulse width saturation method, (c) relative comparison using a high accuracy Multi pixel photon counter (MPPC) coupled scintillator gamma radiation detector, and (d) gross method are used. We tested above methods to find out the best / optimal option to calibrate the electronic setting of the detector and decide for other parameters (like time and number of samples) to measure counts with better accuracy and precision. It is also tested (for the given data set) which one, mean, median or mode would be an optimal statistical tool for the given data set. Finally, we test, which of the electronics settings are more important to pay attention to before one starts the measurement. Brief description about methods, our methodology to test them, are discussed followed by results.

## 2. Theory:

Statistical tools based on saturation in repeating values and sharp probability distribution are used in this work. The central limit theorem states that for a sequence of a random independent variable having some finite variance then the mean of the sequence tends towards Gaussian / normal distribution as the number of variables goes larger (Trotter, 1959)(Brosamler, 1988). Two methods based on the central limit theorem are used to fit the experimental data into the Gaussian model are curve fitting (Gauch Jr and Chase, 1974) and Gaussian probability distribution(Xu et al., 2018). The Gaussian function has been widely used in modeling and to explain the behavior of physical phenomena in various disciplines of applied physics and engineering. For example, the Gaussian function has been used to model practical microscopic applications(Zhang et al., 2007), LASER heat source in LASER transmission welding(Liu et al., 2016). In the field of signal and image processing Gaussian function is used to model the approximation of Airy disk(Guo, 2011). In applied science, the random noise that corrupts the signal can be modeled with the Gaussian function according to the central limit theorem(Al-Nahhal et al., 2019). Following are a brief description:

### 2.1. Optimal value calculation:

**Method 1 (CLT1):** - This method uses Gaussian curve fitting with the following equation (Eq. 2):

$$C_{x_i^j}(M) = A_i^j e^{-\left(\frac{\left(M-b_i^j\right)}{c_i^j}\right)} + D_i^j e^{-\left(\frac{\left(M-e_i^j\right)}{f_i^j}\right)} \quad \text{where } i = 1,2,\ldots,N = 5; j = 1,2,3,\ldots r. \quad (2)$$

To find the optimal setting/value of a particular parameter $x_i^j$, it is varied within its set range (for example recording time is varied between $r = 5$ till 30sec) while keeping other parameter values at a certain constant value. For each variation, the number of radiation counts $C_{x_i^j}$ is recorded. This is repeated 30 times ($M$). The radiation counts $C_{x_i^j}(M)$ with regards to instance $M$ are fitted according to Gaussian. Each plot is fitted into shape controlling parameters, A, $b$, $c, D, e$, and $f$. The corresponding value of $x_i^j$ with the plot having best-fitting parameters with minimum sum of squared deviation and higher goodness of fit is termed as optimal value $x_i^{opti.}$. One example is shown in Fig. 1a. It is shown that radiation count (normalized w.r.t maximum value) plots vary (shape and amplitude), if, the same experiment is repeated multiple times using a different combination of parameters. The one which is having closeness with Gaussian is termed as an optimal combination.

**Method 2 (CLT2):** - This method declares the value $x_i^j$ as optimal $x_i^{opti.}$, if, its probability distribution function plot with regards to instances has higher probability values, minimum full width half maxima (FWHM), high peak amplitude, and low standard deviation values. The definition of the normal probability density function is given in the following equation (Eq. 3):

$$p_{x_i^j} = \frac{1}{\sigma_i^j \sqrt{2\pi}} e^{\left(\frac{-\left(c_{i,M}^j - \bar{c}_{i,M}^j\right)^2}{2(\sigma_i^j)^2}\right)} \quad \text{where } i = 1,2,\ldots,N = 5; j = 1,2,3,\ldots r. \quad (3)$$

Where $\sigma$ is standard deviation and $\bar{x}_{i,M}^{j}$ is the location parameter (it could be mean, mode or median) of the distribution depending on the respective case study. Probability distribution $p_{x_i^j}$ of radiation counts $c$ with regards to instance $M$ is plotted for each variation (Johnson, N. L., S. Kotz, n.d.). One example is shown in Fig. 1b. The probability distribution function for four data (obtained by setting different combinations of LLD, Voltage, M, T, and Amp. Gain values) are plotted against instances $M$ (30 values interpolated into 1500 values for better visualization). The combination with the highest peak value with the smallest full width at half max is termed as optimal.

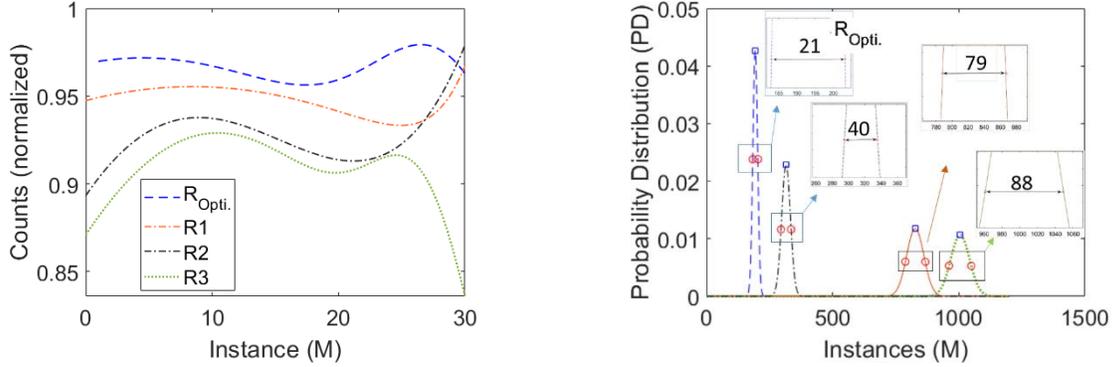

(a) Method 1: count (normalized w.r.t. max val.) vs. Instance plot; optimal combination ($R_{Opti.}$) and suboptimal combinations ($R1, R2, and\ R3$) of parameter setting are estimated by Gaussian Fits.

(b) Method 2: Probability Distribution (PD) vs. Instances (M) (30 values are interpolated into 1500); Center Limit theorem showing optimal combination having the sharpest probability distribution.

Figure 1. Estimation of Optimal combination by Method 1 and Method 2.

**Method 3 (PWS): -** This method chooses optimality based on most flat / plateau/saturation region assuming that the number of counts is less dependent on the setting of parameters. The combination of parameters setting for which flat region is obtained is considered as the optimal setting of the detector.

**Method 4 (True counts measurement): -** Counts measured for all combinations of parameter variation measured by Na(Tl) based scintillation detector are compared with the counts obtained by MPPC coupled scintillator detector("Radiation detection module C12137 | Hamamatsu Photonics," n.d.). The corresponding combination of parameters with the least absolute difference is assumed optimal explained in Eq. 4. This method assumes that counts measured using MPPC coupled scintillator detector $\bar{c}_{T\,i,M}^{\,j}$ are near to true radiation counts.

$$x_i^{opti.} = \bar{x}_{i,M}^{j} \big| \min(\big| \bar{c}_{i,M}^{j} - \bar{c}_{T\,i,M}^{\,j} \big|)$$

(4)

**Method 5 (Gross Method):-** According to the definition of this approach, the optimal parameter value would have the highest ratio of the square of gross sample counts rate to background counts rate $\bar{c}_0$. Its relation is given in Eq. 5:

$$x_i^{opti.} = \bar{x}_{i,M}^{j} \Big| \max\Big(\frac{\bar{c}_{i,M}^{j\,2}}{\bar{c}_{0\,i,M}^{\,j}}\Big)$$

(5)

## 3. Methodology:

The exhaustive brute force procedure is adopted to find out the optimal combination of four parameters (Time, LLD, amplifier gain, and high voltage applied at PMT) set by users to detect counts. Fig.2 shows the flow of the work.

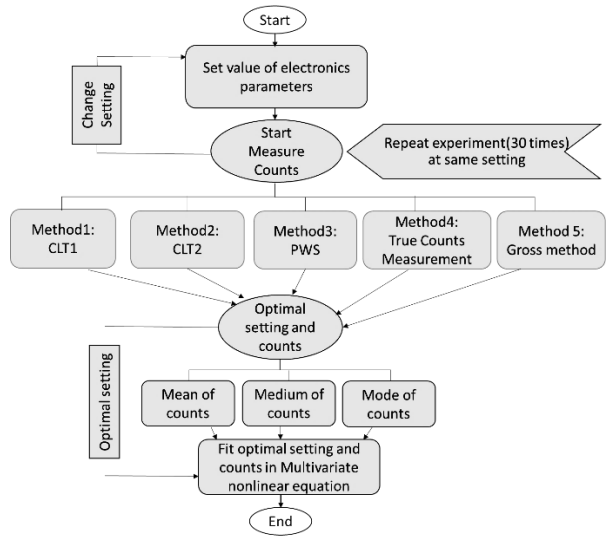

Figure 2. Flow chart of the adopted methodology

Gamma radiation source Cesium 137 ($^{137}Cs$) of 1.5 µCi and a total of 6 scintillation detectors and associated electronics of two different make/manufacturer is used. Five probe geometry NaI(Tl) scintillator detectors (a crystal of diameter 1 inch) are arranged in fan-beam to include direction dependency (*Model PNS-2 Para Nuclear Spectrometer*, n.d., "Para Electronics - Home," n.d.). The measured data is transferred to the computer system using Nuclear Gamma Spectrometer PNS-2™ software, semi-automatically. The experimental setup is shown in Fig.3. The operational voltage rating $V_{cal.}$ (estimated onsite on the date of delivery) and amplifier gain of each scintillator detector is provided by the Manufacturer. Details are mentioned in Table 1. Another scintillator detector (CsI(Tl) crystal of size: 13 x 13 x 20 mm) coupled with multi-pixel photon counters photomultiplier (MPPC) is used for confirming the readings assuming this detector counts the radiation resembles true counts. Cathode to anode voltage at PMT, gain at the amplifier, window level of SCA or lower-level discriminator (LLD), Time to count events are assumed to be independent variables. The Brute Force approach is used by varying each of these parameters in apt ranges, one at a time. The ranges of every parameter is given as- Time: 05sec to 30sec with resolution of 5sec, Voltage: 600V to 900V with resolution 50 V, Amplifier: 0.4 to 2.8 with resolution 0.4 and LLD: 0.4 to 1.6 with resolution 0.2. The experiment was carried out at every combination of four independent parameters and consequently, we get 2058 combinations of data set for every single detector. The experiment was repeated 30 times at every combination of settings. The same procedure is repeated for the rest of the detectors.

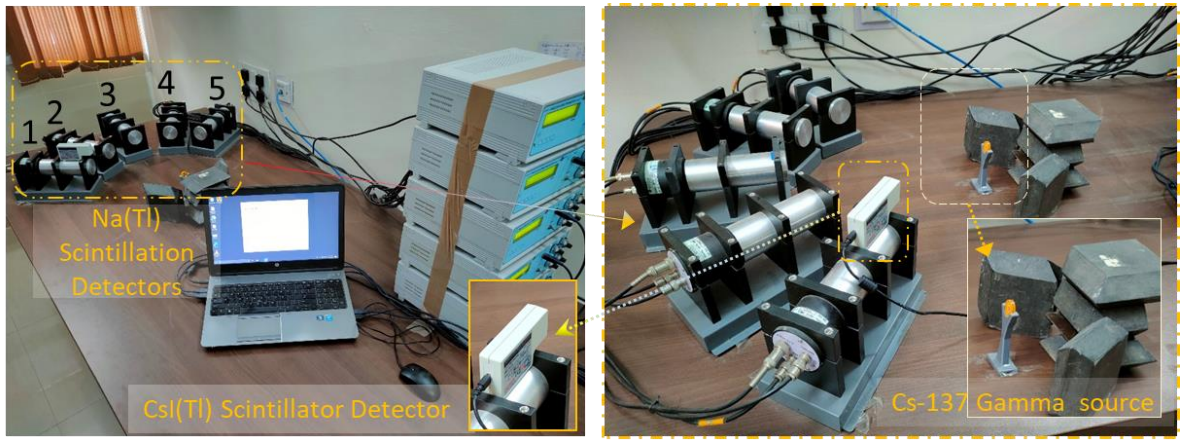

Fig. 3: Experimental setup for measuring Counts using scintillation detector.

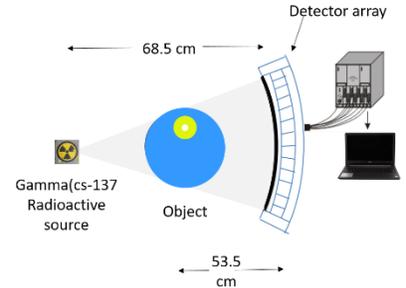

Gamma CT is used to see the impact of calibration of the detector over the accuracy of the result. The geometry of the gamma CT system is shown in Fig. 4. Measurement was performed using a phantom, object with known inner profile (shown in Fig. 4) at two different settings: (a) one at the optimal setting calculated by the best method and (b) the other at a random setting. Phantom used in our Gamma CT experiment consists of three different materials: Perspex ( μ~1.02), aluminum ( μ~0.211), and steel (μ~0.544) where μ is the attenuation coefficient ($cm^{-1}$) of material (Goswami et al., 2016). Distance between source and detector is kept 68.5 cm, object to the detector is 53.5, fan beam angle was 47.58 cm. and diameter of the object is 12 cm. Data acquired at every rotation angle $20°$ up to a complete rotation of $360°$.

Fig. 4: Geometry of gamma CT system

### 3.1. Choice of best method and Parameters with major contribution:

Optimal values of LLD, time, amplifier, and voltage of each detector are estimated using 5 different methods. In data processing steps, when a single value is required in few methods (method 4 and 5), three different case studies/results are created by taking the mean, mode, and median of the same data and estimating the optimal values of parameters. This way we incorporated and studying optimal settings having influenced from these statistical tools. Counts measured at optimal setting from each method are compared with count rate of the radioactive source. CT measurements (for median, mean and mode of data) are made for settings estimated from all 5 calibration methods. It is assumed that settings / parameters estimated using optimal calibration method would give best reconstruction and would have similar count rates.

Optimal settings of all detectors, calculated by a particular method, are fitted into the multivariate nonlinear equation (Eq. 6) to see the major contribution of the parameter. Inbuilt function 'nlinfit' of MATLAB® is used for nonlinear fitting (Dumouchel and O'Brien, 1989; Holland and Welsch, 1977; Seber and Wild, 2003). Robust fitting with weight function 'Talwar' is used (Hinich and Talwar, 1975). We note that when the number of outcomes is too small in multivariate nonlinear regression results from the analysis may not be accurate and precise (Peduzzi et al., 1995). We choose two different models. The other one has one more independent parameter instance/sample size ($M$) to testify its influence over measurement results. Following nonlinear polynomial equations (Eq.6 and 7) are used for empirical fitting the data:

$$counts = b_1(LLD)^{b2} + b_3(Time)^{b4} + b_5(Amp)^{b6} + b_7(Volt)^{b8} + b_9 \qquad (6)$$

$$counts = b_1(LLD)^{b2} + b_3(Time)^{b4} + b_5(Amp)^{b6} + b_7(Volt)^{b8} + b_9(M)^{b10} + b_{11} \qquad (7)$$

To avoid falling into a local minimum, we ran the codes for several fine boundary conditions of coefficients ($b_1$ to $b_9$) or $b_{11}$ in between 0.1 to 7.1 with 0.2 resolution. Maximum entropy and minimum $l_2$ norm are used as an objective function in the above multivariate nonlinear constraint optimization analysis (Chai and Draxler, 2014);(Hui and Liu, 2020));(Barron, 1986).

It is assumed that mutually dependent higher-order terms would have negligible contribution as the parameters are considered independent to each other. As a final check, another practical constraint based on human perception is applied. Out of many solutions with very close values, we chose the one that has positive values of $b_3$ and $b_7$ as the time and voltage would increase counts will increase. The optimizing results having large contribution of constant coefficient $b_9$ or $b_{11}$ are rejected to ensure contribution from other parameters / settings. This optimized nonlinear equation is used to find the prominent term having larger contribution to effect the radiation measurement. In next section we now discuss the results.

## 4. Result and discussion:

Detectors are calibrated using five different methods briefly discussed in the theory section. Optimal values of amplifier gain, LLD, high voltage at PMT, Time to measure counts for all detectors are shown in table 2. The respective counts are shown in fig. 5. Figure 5 (a), (b), (c), (d), and (e) shows the counts detected by 5 detectors at optimal setting calculated by method1, method 2, method 3, method 4, and method 5 respectively. It is observed that optimal counts calculated by method 2 (CLT2) are close to counts rate of radioactive source. Multivariate nonlinear equation eq. 6 modeled to fit optimal setting parameters and corresponding counts calculated by all methods.

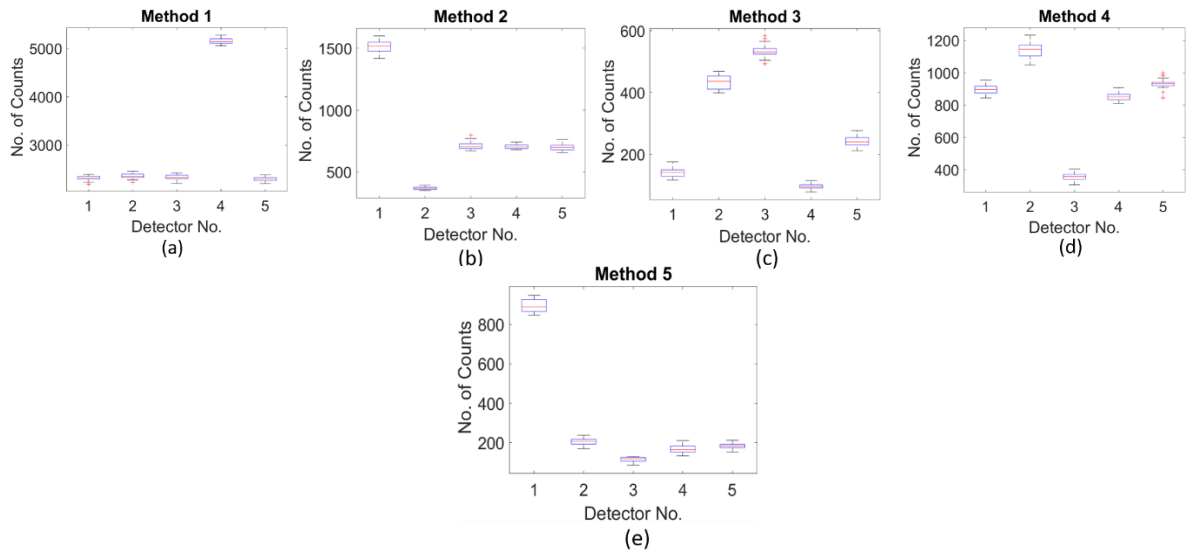

Fig 5. Optimal counts Calculated by (a) method 1 (CLT1), (b) method 2 (CLT2), (c) method 3 (PWS), (d) method 4 (True counts measurement), (e) method 5 (Gross method).

Table 2: Optimal setting of detectors calculated by different methods.

| Method 1 | Sample 30 | Det. 1 | Det. 2 | Det. 3 | Det. 4 | Det. 4 |
|---|---|---|---|---|---|---|
| | LLD | 0.8 | 0.6 | 0.6 | 0.6 | 0.6 |
| | Time(sec) | 25 | 30 | 30 | 30 | 20 |
| | Amp. Gain | 2.8 | 0.4 | 1.6 | 2.0 | 2.8 |
| | Voltage (V) | 601 | 701 | 700 | 601 | 900 |
| Method 2 | LLD | 1.2 | 1.2 | 1.2 | 0.6 | 0.8 |
| | Time(sec) | 25 | 30 | 10 | 10 | 15 |
| | Amp. Gain | 0.8 | 2.0 | 2.8 | 0.8 | 1.6 |
| | Voltage (V) | 750 | 750 | 601 | 651 | 750 |
| Method 3 | LLD | 0.4 | 0.4 | 0.4 | 1.4 | 0.4 |
| | Time(sec) | 5 | 5 | 5 | 5 | 5 |
| | Amp. Gain | 2.8 | 2.8 | 2.0 | 2.4 | 2.4 |
| | Voltage (V) | 750 | 750 | 750 | 900 | 750 |
| Method4 | LLD | 1.0 | 0.8 | 1.0 | 1.0 | 1.0 |
| | Time(sec) | 15 | 15 | 25 | 25 | 25 |
| | Amp. Gain | 2.8 | 0.4 | 0.8 | 1.6 | 1.2 |
| | Voltage (V) | 651 | 900 | 600 | 850 | 651 |
| Method 5 | LLD | 0.4 | 0.6 | 0.6 | 0.4 | 0.6 |
| | Time(sec) | 5 | 5 | 5 | 5 | 5 |
| | Amp. Gain | 0.4 | 2.0 | 0.4 | 0.4 | 0.4 |
| | Voltage (V) | 651 | 651 | 750 | 701 | 750 |

Entropy, coefficients of time ($b_3$, $b_4$), coefficients of voltage ($b_7$, $b_8$) and constant parameter ($b_9$) of 5 different methods are plotted in figures 6 (a), 6(b), 6(c), 6(d), 6(e) and 6(f) respectively. Data are plotted for three statistical tools mean, median and mode. In fig. 6(a) we see that all methods have comparatively equal value of entropy except for method 5 irrespective of mean, median, and mode. In figure 6(d) we got a very large value of $b_9$ coefficient for method 3 so we divided this parameter by appropriate value so it will not suppress the value of the coefficient for other methods for easy comparison. It can be observed from figs. 6(b), 6(c), 6(d), and 6(e) that methods 1, 3, and 4 are dropped out from further discussion due to practical constraints (positive coefficient for voltage and time). We get almost near to zero error values hence no comparison can be done on basis of the plot for methods 1, 2, and 4. We observed a large value of root mean square error (RMSE) for methods 3 and 5 compared to other methods which show these methods are not fitted well in our modeled equation. The constant term / coefficient $b_9$ is small as well for method2 for mean (least), median and mode cases. Entropy values are also consistently comparable with maximum values for method2. Figure 6 concludes that method 2 is fitted best in the multivariate nonlinear equation and having optimal counts close to counts rate of radioactive source. Over all discussion leads to mean as a recommended statistical parameter that preserves accuracy and precision of the measurement.

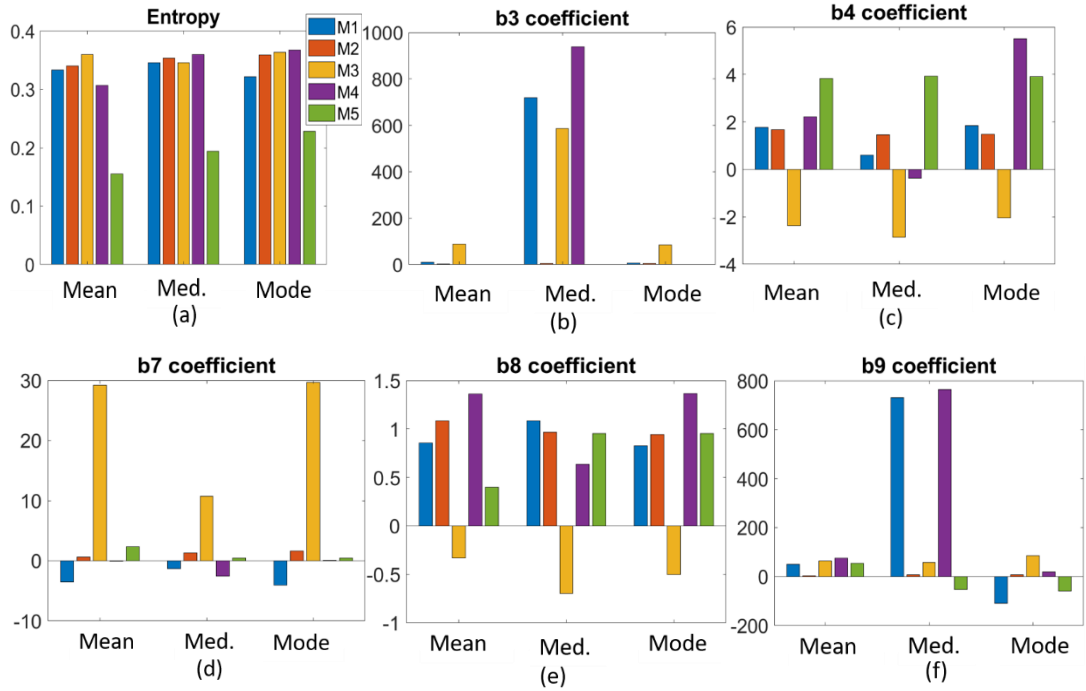

**Fig. 6:** For 5 different methods and their mean, median, and mode (a) entropy, (b) time multiplied coefficient, (c) time exponential coefficient, (d) voltage multiplied coefficient, (e) voltage exponential coefficient, and (f) $b_9$ coefficient.

The optimized nonlinear eq. for four independent parameters is:

$$Counts = (-0.125) * (LLD)^{(-13.50)} + (2.4) * (Time)^{1.67} + (-3.09) * (Amp)^{4.93} + (0.63) * (Volt)^{1.09} + 3.34 \tag{8}$$

Sample number ($M$) is considered as an independent variable to see an impact on the statistics of the measurement in next stage of analysis. Up to sample number 25, we got a negative coefficient of sample number. Sample size coefficient negative tells that this is not the threshold sample size. The threshold value of the sample number is observed at 26 in our work. We varied sample numbers from 26 to 30 with resolution 1.

Redoing the above analysis and including $M$ as independent variable again resulted method2 (CLT2) as best option. Optimal counts and settings are fitted in modeled equation, eq. 7, and observed that sample number 30 yields the best result, best in the sense it satisfies the criteria that we decided to choose optimal equation. Entropy and coefficient ($b_{11}$) of equation 7 are plotted in figs. 7(a) and (b) respectively for mean, median, and mode of counts. In this case, we observed that the median optimizes the result. We found that the data has very negligible skewness and plots are skipped here for maintaining the brevity. The optimal nonlinear equation for five independent parameters is:

$$Counts = -0.397 * (LLD)^{-11.72} + (5.18) * (Time)^{1.46} + (-1.74) * (Amp)^{5.60} + (4.55) * (Volt)^{0.75} + (3.01) * (M)^{1.07} + 5.36 \tag{9}$$

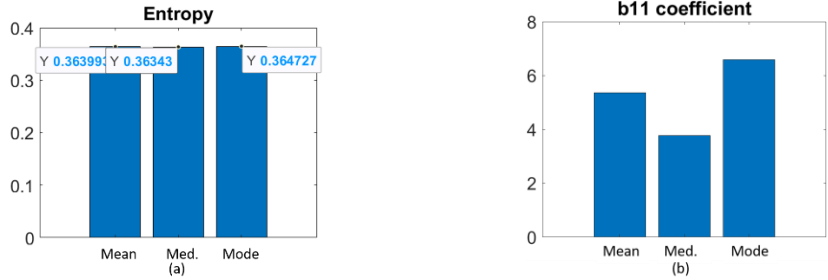

Fig. 7: (a) entropy, (b) $b_{11}$ coefficient for modelled equation 7.

### 4.1. Gamma CT Analysis:

CT measurements are performed for (a) random parameter values, and optimal settings obtained from all methods. Reconstruction results are shown in Fig.8 for random setting and optimal setting obtained from Method2. Upper bottom in Fig. 8 shows (a) real picture of phantom, (b) cyber phantom: 3-D image of phantom, and (c) its digital analogues. Figure 8(d) shows the reconstructed image of cyber phantom obtained by (d) projection data created using digital analogous / simulation with 33% RMSE. Figure 8(e) shows reconstruction from real data acquired at random setting and fig. 8(f) at optimal setting calculated by best method 2 (CLT2). Hybrid reconstruction method, MaxenT is used to obtain reconstructed images (Goswami et al., 2015). Reconstruction error is minimum for real data measurement when optimal calibration setting is used. Root mean square error is calculated with respect to cyber phantom image (Fig. 8c).

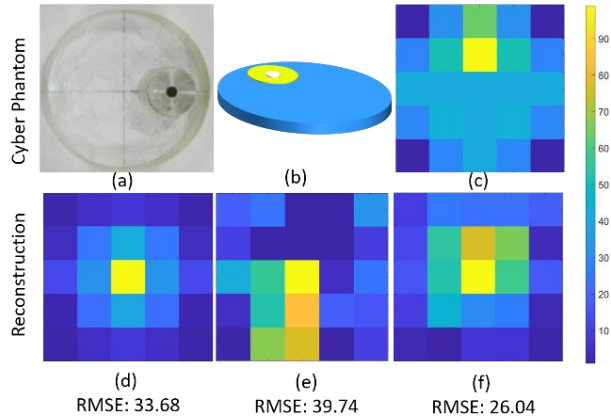

Fig. 8 (a) camera image of phantom, (b) 3-D image, (c) cyber phantom, (d) simulation, (e) at random setting, (f) at optimal setting. The reconstruction with optimal settings has least RMSE and better visual resemblance with true.

Scattering is included in all of the measurement as shielding is not used and optimal calibration setting may change. The optimal nonlinear equation gives a relationship between counts and the optimal value of LLD, time, Amplifier gain, and high voltage. Both optimal nonlinear equation for 4 and 5 independent parameters shows that time to measure the radiation count has a dominant influence over other.

Each detector has different optimal calibration settings owning to the fact that existing fabrication techniques result (structure: electronics, impurities of Thallium in Na) different detectors. Direction dependency may be involved and further study is required.

Despite all detectors are not the same, we are forcing the optimal setting values from all detectors to curve fit Eq. 6 & 7. It is assumed that this would give us bulk characteristics for a major contribution of the parameter.

Sparse data set is used to develop equation 6. Eq. 6 & 7 are made of neglecting several mutual dependent terms, assuming the parameters are independent of each other.

## 5. Conclusion

Following are the pointwise conclusions:

1. Median is found to be best statistical tool in this work, if sample size is include as independent parameter during the calibration stage. This observation is as per according to previously reported works. Otherwise, Mean as a statistical tool is recommended. Data was found to have negligible skewness limiting the scope to provide comments about utility of mode and median.
2. Second version of Center Limiting Theorem has found to be optimal calibration method.
3. The optimal calibration settings corresponds to better practical feasibility for precision and accuracy as CT reconstruction found to be best.
4. Multivariate nonlinear analysis suggest that once the LLD is set (according to photopeak at center), it is essential that time to measure the radiation counts must be optimally obtained. Any random duration for measurement may affect the results. Voltage and sample size are other important parameter to look for. A strategy is presented here for calibration.

Several assumptions are made in this work, however are under reasonable limit. Similar analysis is performed using conventional Nuclear Radiation bins with manual recording interface. The proposed strategy is universal/general as it can be applied to any other field of radiation measurement.

### Author contribution:
Kajal has contributed in experiments, analysis. Mayank has designed the study, contributed in analysis, coding, writing text and arranging funds for resources.

### Acknowledgement:
This work is partially supported by the Department of Science and Technology-Science and Engineering Research Board (DST-SERB) under Early Career scheme, Project number: ECR/2017/001432, Government of India and Office of Dean Finance & Planning, IIT Roorkee, Roorkee, India. We acknowledge Mr. Kumar Abinash Misra, DIC Intern from Dept. of Mechanical Engineering, IIT Mandi for helping us to take initial data.

### Competing Interests:
The authors declare that no competing interest exists.